\documentclass{PoS}
\usepackage{graphicx}
\def\lsim{\mathrel{\rlap{\lower3pt\hbox{\hskip1pt$\sim$}}
\raise1pt\hbox{$<$}}}
\def\gsim{\mathrel{\rlap{\lower3pt\hbox{\hskip1pt$\sim$}}
\raise1pt\hbox{$>$}}}

\newcommand{\beq}{\begin{equation}}
\newcommand{\eeq}{\end{equation}}
\newcommand{\bea}{\begin{eqnarray}}
\newcommand{\eea}{\end{eqnarray}}

\title{Recent results on the Equation of State of QCD}
\ShortTitle{EoS of QCD}
\author{
Szabolcs~Bors\'{a}nyi$^a$,
Zolt\'{a}n~Fodor$^{a,b,c}$,
Christian~Hoelbling$^{a}$, 
S\'{a}ndor~D.~Katz$^{c,d}$,
\speaker{Stefan Krieg}$^{,a,b}$,
Claudia Ratti$^{e}$,
K\'alm\'an~K.~Szab\'o$^{a,b}$\\
\ \\
$^a$Department of Physics, University of Wuppertal,Gau\ss str. 20, D-42119, 
Germany\\
$^b$Forschungszentrum J\"ulich, J\"ulich, D-52425, Germany\\
$^c$Institute for Theoretical Physics, E\"otv\"os University, P\'azm\'any
1, H-1117 Budapest, Hungary\\
$^d$MTA-ELTE Lend\"ulet Lattice Gauge Theory Research Group\\
$^e$Università di Torino and INFN, Sezione di Torino via Giuria 1,
 I-10125 Torino, Italy
}
\FullConference{The 32nd International Symposium on Lattice Field Theory,\\
		23-28 June, 2014\\
		Columbia University New York, NY}
\abstract{
We report on a continuum extrapolated result~\cite{Borsanyi:2013bia} for the equation of state (EoS) of QCD with $N_f=2+1$ dynamical quark flavors and discuss preliminary results obtained with an additional dynamical charm quark ($N_f=2+1+1$). For all our final results, the systematics are controlled, quark masses are set to their physical values, and the continuum limit is taken using at least three lattice spacings corresponding to temporal extents up to $N_t=16$. 
}

\begin{document}

\section{Full result for the $N_f=2+1$ equation of state}
The rapid transition from the quark-gluon-plasma 'phase'\footnote{Since this transition is a cross-over~\cite{Aoki:2006we}, this use of the term 'phase' is somewhat abusive, and indicates only the dominant degrees of freedom.} to the hadronic phase in the early universe and the QCD phase diagram are subjects of intense study in present heavy-ion experiments (LHC@CERN, RHIC@BNL, and the upcoming FAIR@GSI). This transition can be studied in a systematic way in Lattice QCD (for recent reviews see, e.g., \cite{Fodor:2009ax,Philipsen:2012nu,Petreczky:2013qj,Hoelbling:2014uea}).
The equation of state (EoS) of QCD, (i.e, the pressure $p$, energy density $\epsilon$, trace anomaly $I=\epsilon-3p$, entropy $s=(\epsilon+p)/T$, and the speed of sound $c_s^2=dp/d\epsilon$ as functions of the temperature) has been determined by several lattice groups, however, a full result was, until recently, still lacking. Ref.~\cite{Borsanyi:2010cj} constitutes a full result at three characteristic temperatures, which we have now extended to the full temperature range in~\cite{Borsanyi:2013bia} and made available electronically~\cite{resultsonline}.

Our calculation is based on a tree-level Symanzik improved gauge action with 2-step 
stout-link improved staggered fermions. The precise definition of the 
action can be found in ref.~\cite{Aoki:2005vt}, its advantageous scaling properties are studied in ref.~\cite{Aoki:2006br,Aoki:2009sc,Borsanyi:2010bp}. In particular, while it approaches the continuum value of the Stefan-Boltzmann limit in the infinite temperature limit $T\rightarrow\infty$ slower than actions with p4 or Naik terms (the latter is an additional fermionic term in the asqtad and hisq actions), it behaves monotonous and reaches the asymptotic $a^2$ behavior quite early. Extrapolations from moderate temporal 
extents, e.g., using $N_t \ge 8$, allow for a smooth continuum extrapolation and provide an accuracy on the percent level, the typical accuracy one aims to reach. Additionally, applying simple tree-level improvement factors for the bulk thermodynamic observables brings the individual data points for the different $N_t$ very close to the continuum limit. 

\begin{figure}
\begin{center}
\includegraphics*[width=0.470\textwidth]{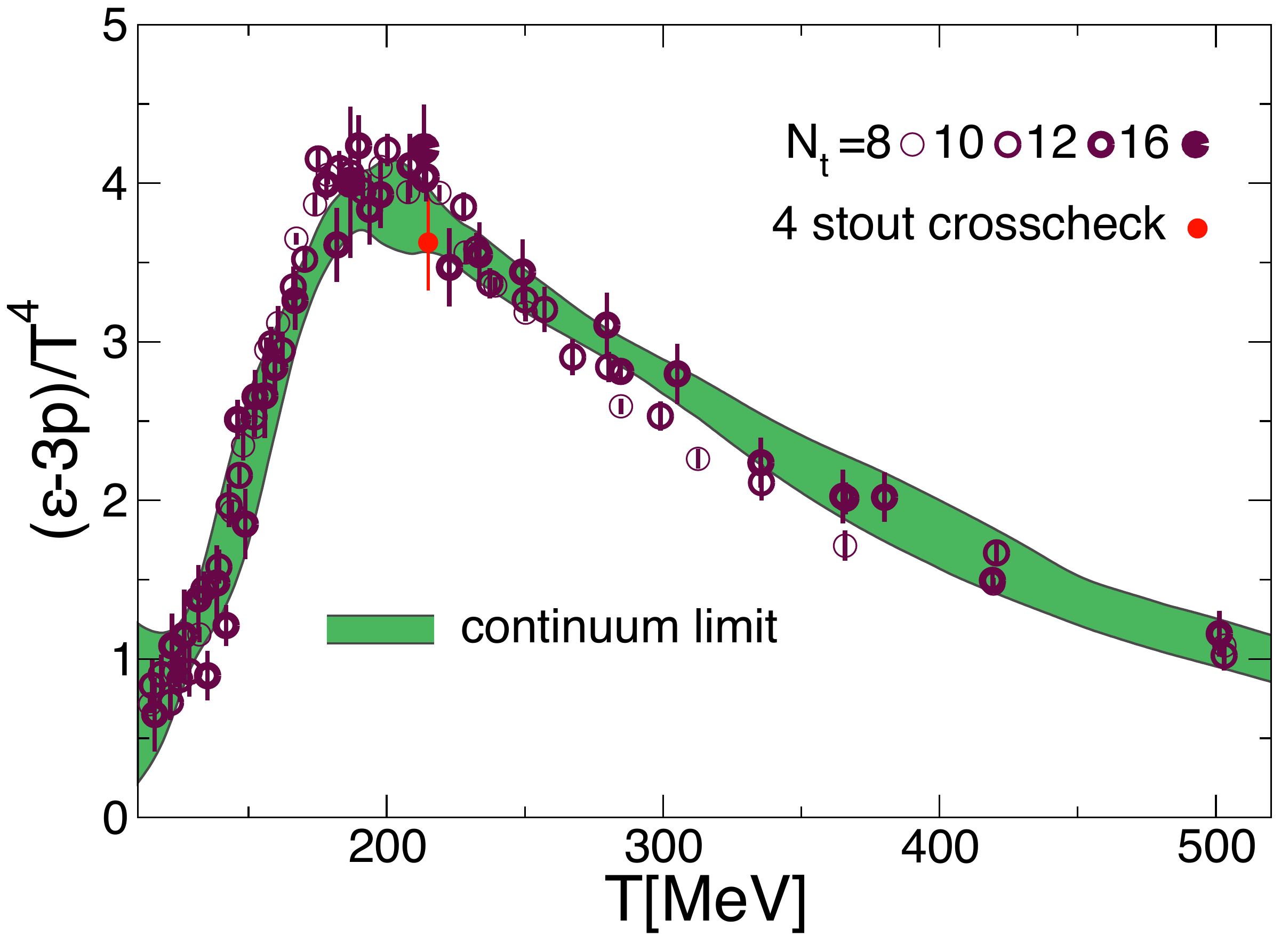}
\includegraphics*[width=0.495\textwidth]{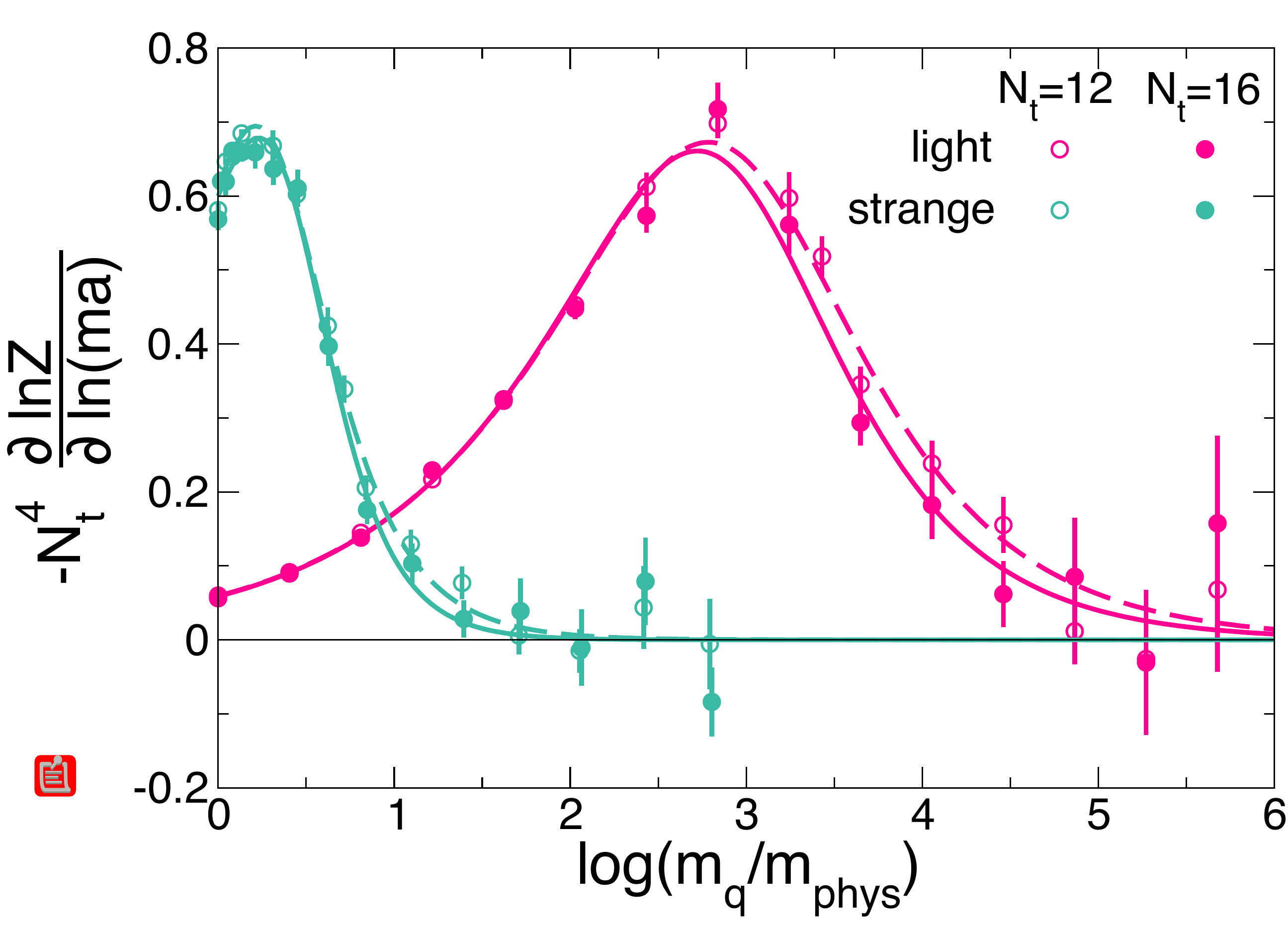}
\end{center}
\caption{\label{tracea_final}
\emph{Left:} The trace anomaly as a function of the temperature. The continuum extrapolated result with total errors is given by the shaded band. Also shown is a cross-check point computed in the continuum limit with our new and different lattice action at $T=214$~MeV, indicated by a smaller filled red point, which serves as a crosscheck on the peak's hight. \emph{Right:} Setting the overall scale of the pressure: integration from the infinitely large mass region down to the physical point using a range of dedicated ensembles and time extents up to $N_t=16$; the sum of the areas under the curves gives $p/T^4$.}
\end{figure}

This points to an important advancement of the calculation described here over our previous results of~\cite{Borsanyi:2010cj}: we now include a large range of $N_t=12$ data points and one $N_t=16$ data point located at the peak position. Previously, we only had $N_t=12$ results at three characteristic temperature values available. The $T\rightarrow0$ limit is difficult due to taste-breaking effects, but is crucial since the renormalization is done at zero temperature, i.e. $p$($T$=0)=0. A mismatch at T=0 leads to a shift in the whole EoS. Previously, we calculated the difference in the pressure between the physical theory and its counterpart with 720~MeV heavy pions at a selected temperature ($100$~MeV), where the latter theory has practically zero pressure, and we, therefore, get $p(T=100~\textrm{MeV})$ in the physical theory with the desired normalization. The difference of this result and the prediction by the Hadron Resonance Gas model (HRG) was then included in the systematical error. With our increased range of temporal extents, we now can use five lattice spacings to fix the additive term in the pressure, integrating in the quark mass down to the physical values (see Figure~\ref{tracea_final}), arriving at a complete agreement with the hadron resonance gas model at low temperatures. We also improved the precision on our line of constant physics (LCP, see ref.~\cite{Borsanyi:2013bia}), and used two different methods to set the scale (based on the $w_0$ scale~\cite{Borsanyi:2012zs} or on $f_k$) in order to control the systematical error related to scale setting. 

These two different scale setting procedures entered into our 'histogram' method~\cite{Durr:2008zz} used to estimate systematical errors, along with a range of other fit methods, each of which is an in principle completely valid approach. We then calculated the goodness of fit Q and weights based on the Akaike information criterion AICc~\cite{AIC,AICc} and looked at the unweighted or weighted (based on Q or AICc) distribution of the results. The median is the central value, whereas the central region containing 68\% of all the possible methods gives an estimate on the systematic uncertainties. This procedure provides very conservative errors. Here, we had four basic types of continuum extrapolation methods (with or without tree level improvement for the pressure and with $a^2$ alone or $a^2$ and $a^4$ discretization effects) and two continuum extrapolation ranges (including or excluding the coarsest lattice $N_t$=6 in the analysis). We used seven ways to determine the subtraction term at T=0 (subtracting directly at the same gauge coupling $\beta$ or interpolating between the $\beta$ values with various orders of interpolation functions), and the aforementioned two scale procedures. Finally, we had eight options to determine the final trace anomaly by choosing among various spline functions, giving altogether 4$\cdot$2$\cdot$7$\cdot$2$\cdot$8=896 methods. Note that using either an AICc or Q based distribution changed the result only by a tiny fraction of the systematic uncertainty. Furthermore, the unweighted distribution always delivered consistent results within systematical errors.

The systematic error procedure clearly demonstrates the robustness of our final result. Even in the case of applying or not applying tree level improvement, where the data points at finite lattice spacing change considerably, the agreement between the continuum extrapolated results, and hence the contribution to the systematic error, is on the few percent level.

The continuum extrapolated trace anomaly is shown in Figure~\ref{tracea_final}.

\section{Update on the $N_f=2+1+1$ Equation of State}
\begin{figure}
\begin{center}
\includegraphics[width=0.41\textwidth]{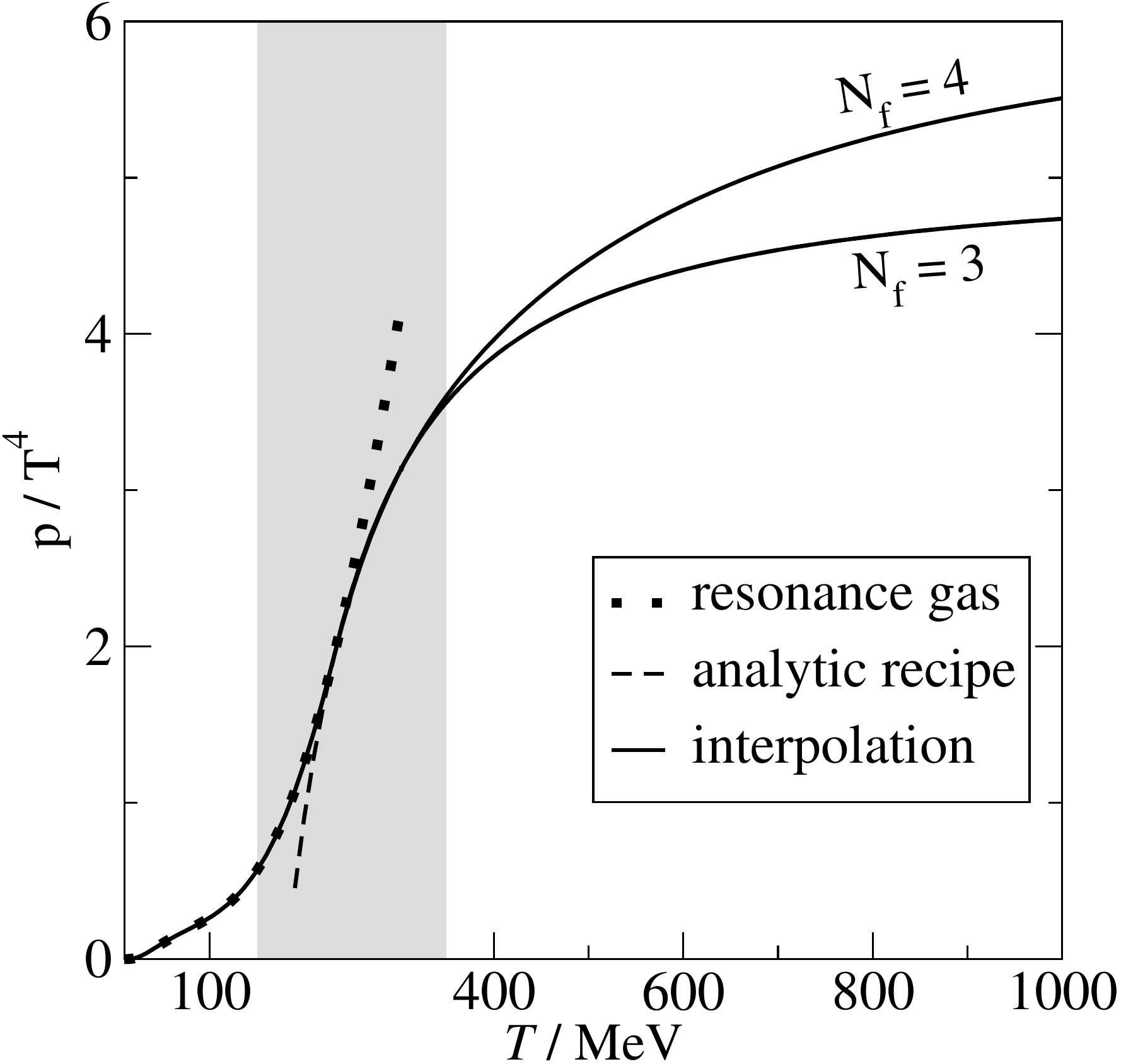}
\includegraphics[width=0.57\textwidth]{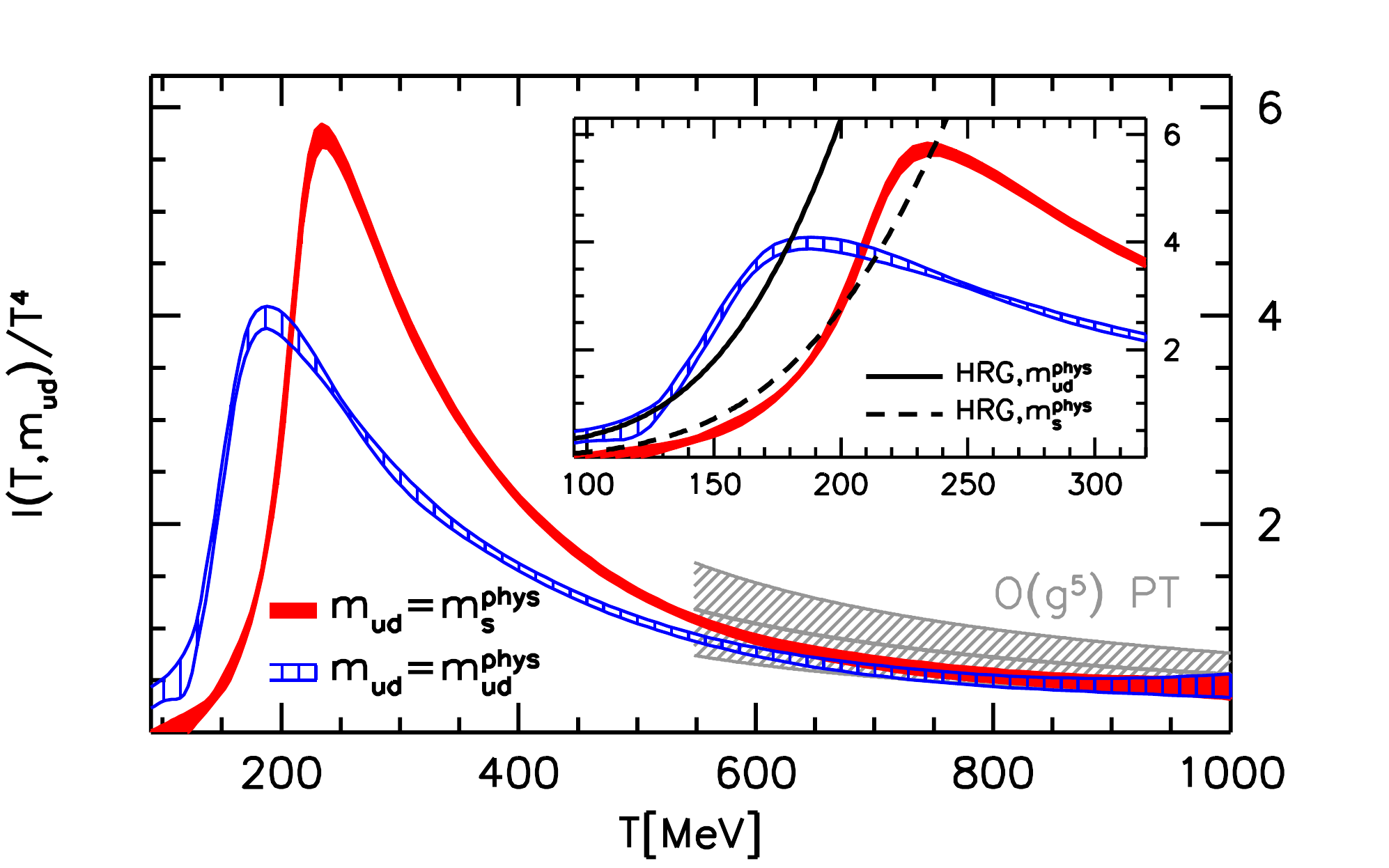}
\end{center}
\caption{\label{fig:ceos}
\emph{Left:}
Laine and Schroeder's perturbative estimate of the effect of the charm in the QCD
equation of state \cite{Laine:2006cp}. \emph{Right:} Wuppertal-Budapest ~\cite{Borsanyi:2010cj}
and perturbative (up to $O(g^5)$) results for the equation of state. 
}
\end{figure}

So far, the equation of state is known only in 2+1 flavor QCD. 
The contribution from the sea charm quarks most
likely matter at least for $T>300-400$~MeV (for an illustration, see Figure~\ref{fig:ceos}).

The $N_f=2+1$ lattice results of the previous section agree with the HRG at low temperatures and are correct for the small to medium temperatures, and, as is shown in Figure~\ref{fig:ceos}, at temperatures of about 1 GeV perturbative results become sufficiently precise. Therefore, we need to calculate the
EoS with a dynamical charm only for the remaining temperatures in the region of approximately $300$~MeV~$<T<1000$~MeV. 

We are using a new lattice action for these calculations. More precisely, where our $N_f=2+1$ calculation used an action with 2 levels of stout gauge link averaging in the coupling of the fermions to the gauge fields, we increased this to 4 levels with a smearing parameter of $\rho=0.125$ (for further details see \cite{Borsanyi:2013bia}). The crosscheck point shown in Figure~\ref{tracea_final} was computed using this new action. Since it perfectly agrees with the $N_f=2+1$ results, even though it was computed using a dynamical charm, we can be certain that at temperatures at and below $T=214$~MeV, we can rely on the $N_f=2+1$ results.

\begin{figure}
\begin{center}
\includegraphics[width=0.49\textwidth]{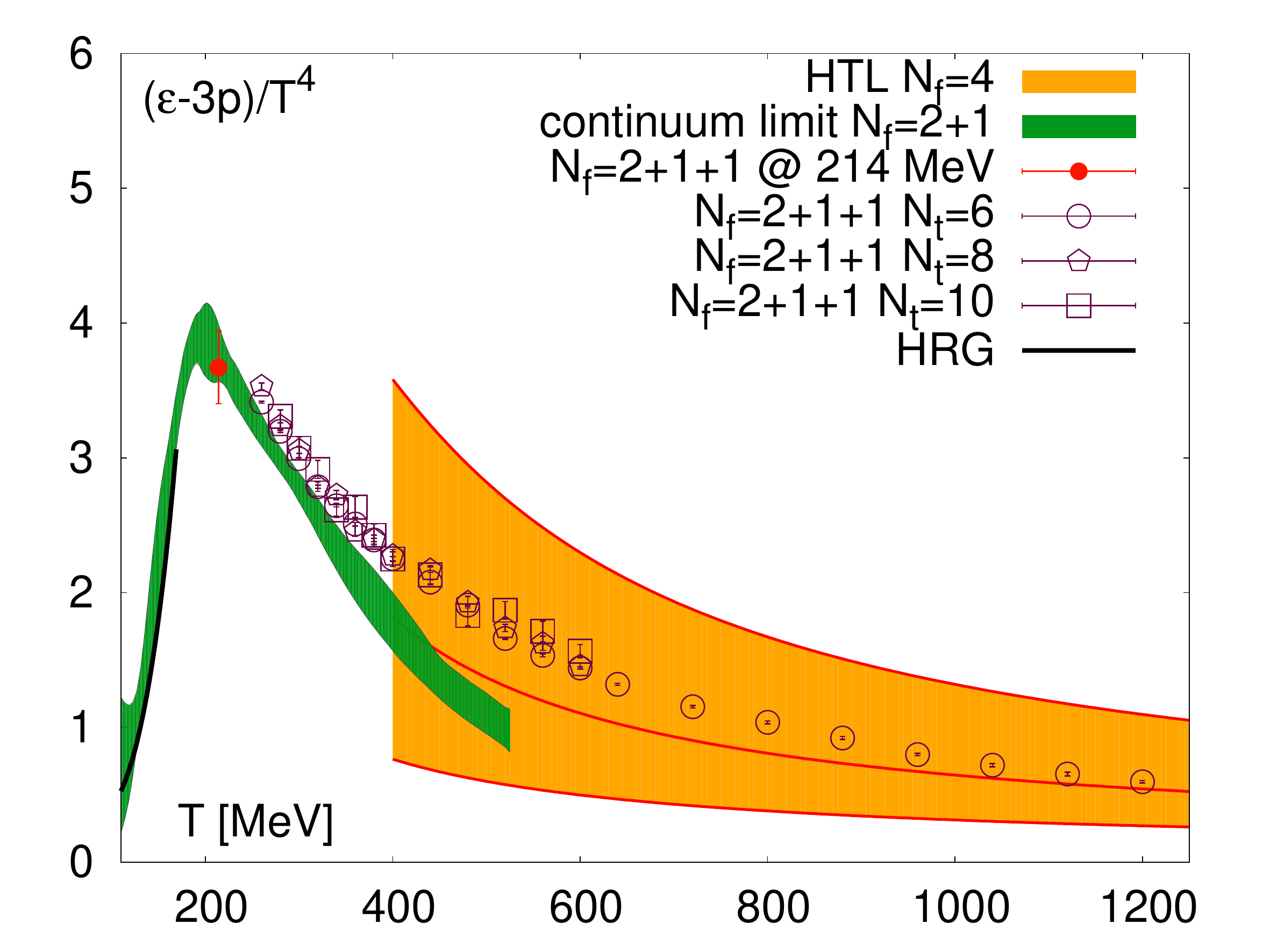}
\includegraphics[width=0.49\textwidth]{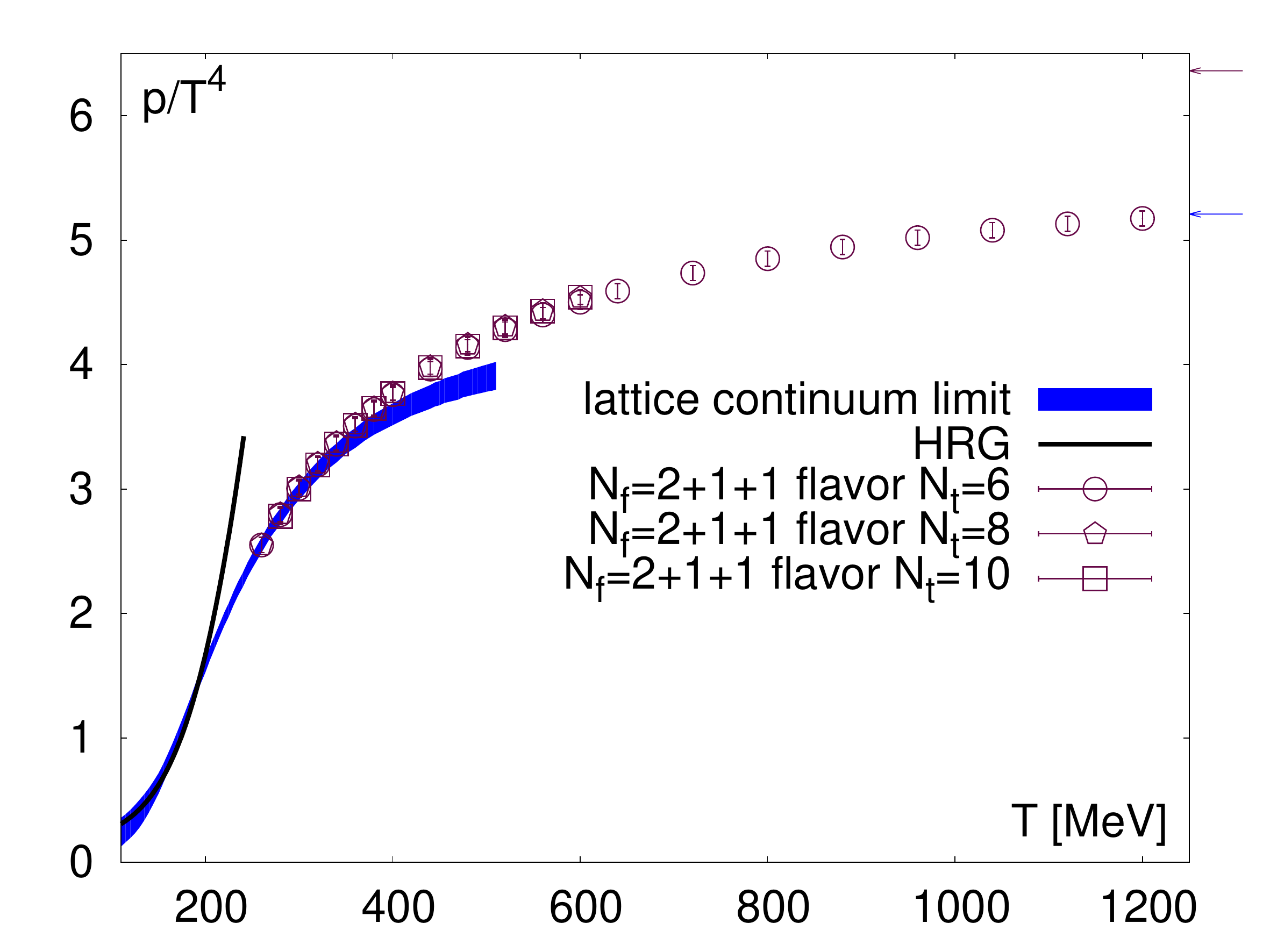}
\end{center}
\caption{\label{fig:dceos}
\emph{Left:} Preliminary results for the charmed EoS. For comparison, we show the HRG result, the $N_f=2+1$ band, and, at high Temperatures, the HTL result~\cite{Andersen:2011sf}, where the central line marks the HTL expectation for the EoS with the band resulting from (large) variations of the renormalization scale.
\emph{Right:} Preliminary result for the pressure, errors indicate the Stefan-Boltzmann value. All errors are statistical only.
}
\end{figure}

Our preliminary results are shown in Figure~\ref{fig:dceos}, all errors are statistical only. Our results span a region of temperatures from $T=214$~MeV up to $T=1.2$~GeV. At the low end we make contact to the $N_f=2+1$ equation of state, and at large temperatures to the HTL result. Thereby, we cover the full region of temperatures, from low temperatures, where the HRG gives reliable results, to high temperatures, where we make contact with perturbation theory. Our present set of data points will be extended to $N_t=12$, in order to make a controlled continuum extrapolation possible.

\subsection{Line of constant physics}
With the switch to a new lattice action comes the need to (re-) compute the LCP. Since we would like to span the temperature range from approximately $300$~MeV~$<T<1000$~MeV, we have to compute the LCP for a large range of couplings or lattice spacings. We split this range up in three overlapping (since we have to make sure that the derivative is smooth) regions according to the applicable simulation strategies.

\begin{figure}
\begin{center}
\includegraphics[width=0.47\textwidth]{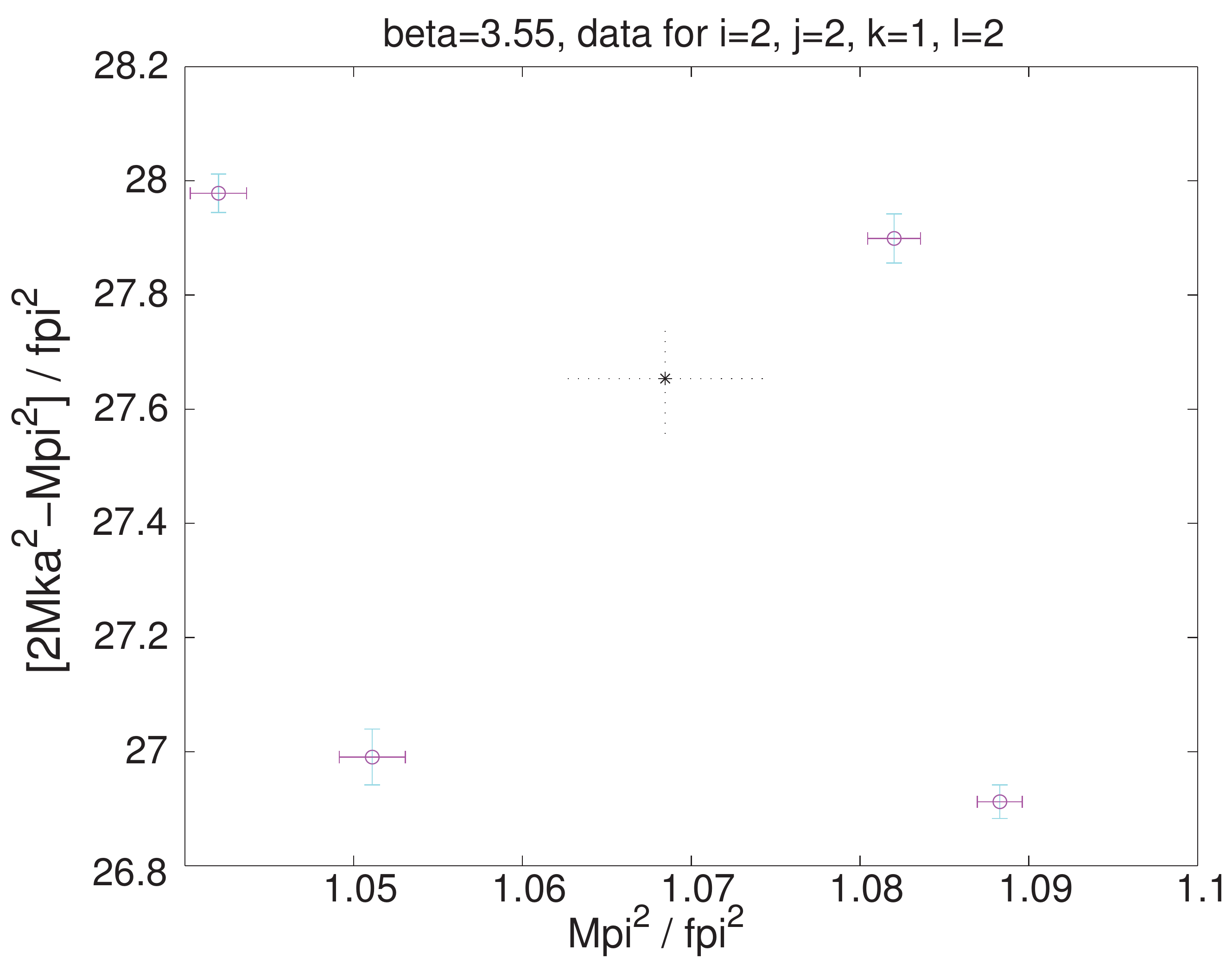}
\includegraphics[width=0.49\textwidth]{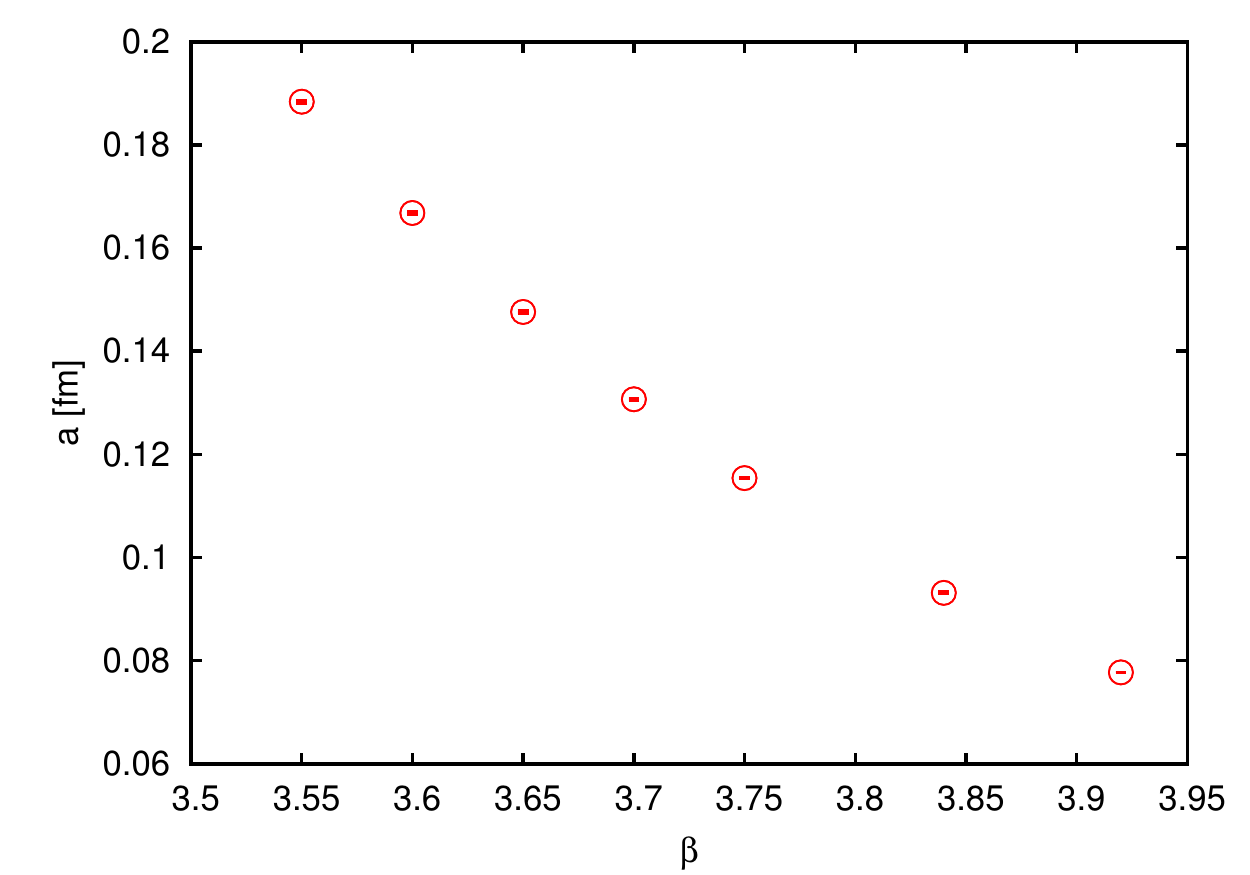}
\end{center}
\caption{\label{fig:lcp-reg1}Region of the LCP, for coarse to medium lattice spacing ($a>0.08$~fm). Here, dedicated simulations bracketing the physical point archive a sub-percent accuracy for the LCP. \emph{Left:} Bracketing of the physical point defined through $M_\pi/f_\pi$ and $(2M_K-M_\pi)/f_\pi$. The strange quark mass is tuned ($m_s/m_l$ is not fixed) and the ratio of the charm to strange quark mass is set at $m_c/m_s=11.85$. \emph{Right:} LCP computed through spectroscopy.
}
\end{figure}

At medium to coarse lattice spacings (region I) one can afford to use spectroscopy to tune the parameters. This is shown in Figure~\ref{fig:lcp-reg1}. Here, we bracketed the physical point defined through $M_\pi/f_\pi$ and $(2M_K-M_\pi)/f_\pi$ and, through interpolation, tune the light and strange quark masses to per-mill precision.

Using the parameters computed in this way, we then performed simulations in the SU(3) flavor-symmetrical point~\cite{Bietenholz:2011qq}, extrapolating the results to our target couplings. There, we tuned the parameters to reproduce the extrapolated results. Since the quark masses are larger than physical, such simulations are considerably less costly than using spectroscopy as for region I, and we are thus able to compute a precise LCP down to fine lattice spacings of $a=0.05$~fm (region II), where the HMC starts being affected by the freezing of topology.  

For finer lattice spacings we thus used our established step scaling procedure \cite{Borsanyi:2013bia} based on the $w_0$ scale. To this end, we computed the observable 
$$
\mathcal{O} = \left.t\frac{d}{dt}\left[ t^2 E(t)\right]\right|_{0.01L^2}
$$
at three different lattice spacings ($a_0$, $a_1$, $a_2$) and volumes ($16^4$, $20^4$, $24^4$) chosen to keep the physical volume fixed, extrapolated to $a_3=24/32 a_2$, and tuned the coupling to match the extrapolated result. Using this method, we extended the LCP to very fine lattice spacings with $a<0.05$~fm (region III).

\begin{figure}
\begin{center}
\includegraphics[width=0.49\textwidth]{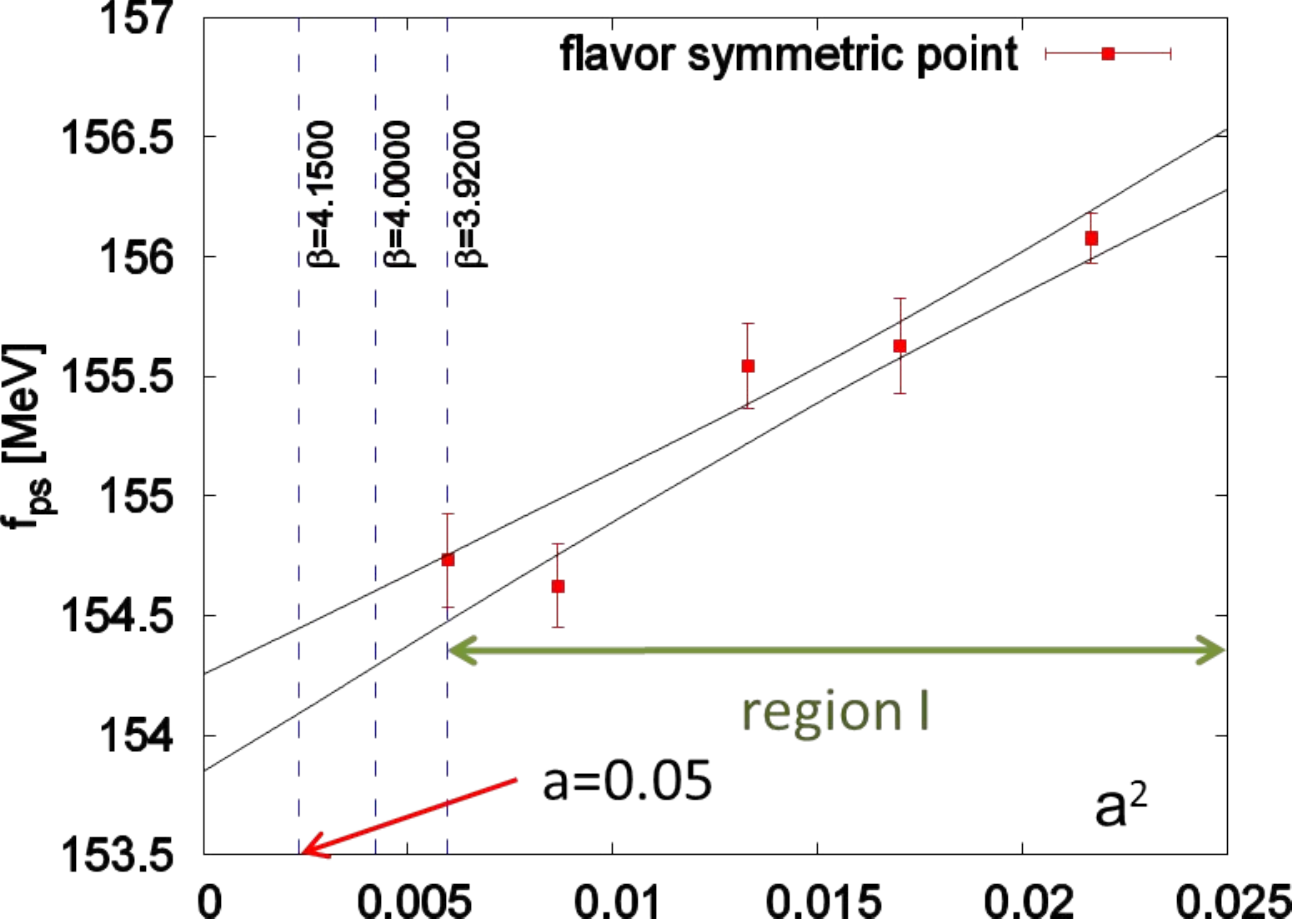}
\includegraphics[width=0.49\textwidth]{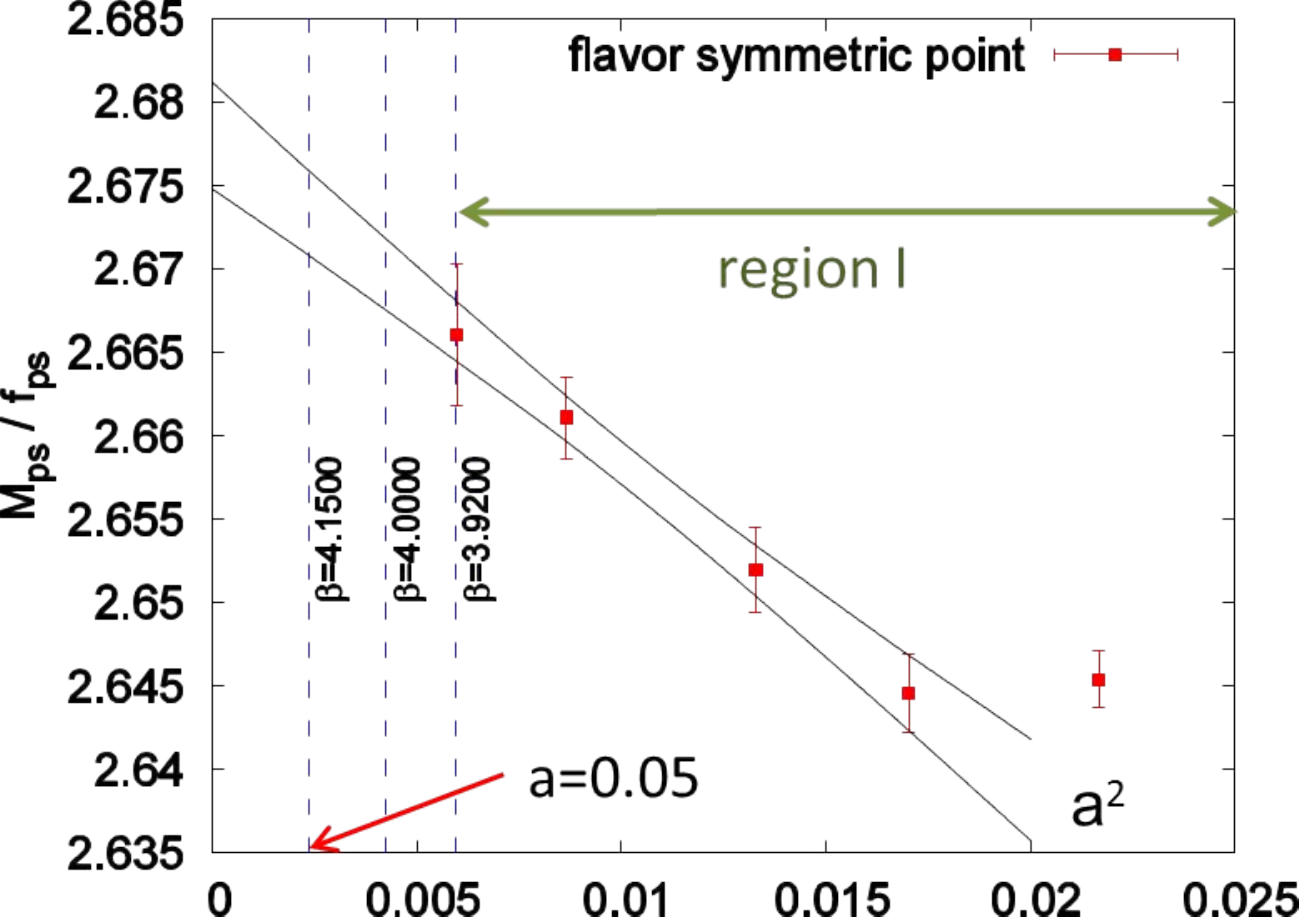}
\end{center}
\caption{\label{fig:lcp2} Using the LCP computed from spectroscopy for coarse to medium lattice spacings (region I), dedicated simulations in the SU(3) flavor-symmetrical point~\cite{Bietenholz:2011qq} using these parameters are extrapolated towards the continuum. At the target coupling, the parameters are tuned until they reproduce the extrapolated value. In this way the LCP is extended to medium to small lattice spacings of $0.08>a>0.05$~fm (region II).
}
\end{figure}

\section{Conclusions}
The precision of Lattice QCD results at finite temperature has increased significantly over the last years. We have discussed a full result (all sources of uncertainties controlled) for the $N_f=2+1$ EoS by the Budapest-Wuppertal collaboration~\cite{Borsanyi:2013bia}, shown how to include a dynamical charm quark for the $N_f=2+1+1$ EoS and presented preliminary results.

\subsection*{Acknowledgments}
Computations were performed on JUQUEEN at Forschungszentrum Jülich, HERMIT at HLRS, Stuttgart, and on the QPACE machine and on GPU clusters~\cite{Egri:2006zm} at University of Wuppertal. We acknowledge PRACE for awarding us resources on JUQUEEN at Forschungszentrum Jülich. This work was partially supported by the DFG Grant SFB/TRR 55 and ERC no. 208740.

\bibliographystyle{utphys}
\bibliography{proceedings}

\end{document}